\documentclass[showpacs,preprintnumbers,floats,prb,a4paper,twocolumn]{revtex4}%
\usepackage{graphicx}
\usepackage{amsmath}
\usepackage{amsfonts}
\usepackage{amssymb}%
\providecommand{\U}[1]{\protect\rule{.1in}{.1in}}
\begin{document}
\preprint{ }
\title{Thermal Instability and Self-Sustained Modulation in Superconducting NbN
Stripline Resonators}
\author{Eran Segev}
\email{segeve@tx.technion.ac.il}
\author{Baleegh Abdo}
\author{Oleg Shtempluck}
\author{Eyal Buks}
\affiliation{Microelectronics Research Center, Department of Electrical Engineering,
Technion, Haifa 32000, Israel}
\date{\today}

\begin{abstract}
We study theoretically and experimentally the response of a microwave
superconducting stripline resonator, integrated with a microbridge, to a
monochromatic injected signal. We find that there is a certain range of
driving parameters, in which a novel nonlinear phenomenon immerges, and
self-sustained modulation of the reflected power off the resonator is
generated by the resonator. A theoretical model which attributes the self
modulation to a thermal instability yields a good agreement with the
experimental results.

\end{abstract}

\maketitle

\section{Introduction}

Nonlinear effects in superconductors have significant implications for both
basic science and technology. Strong nonlinearity may be exploited to study
some important quantum phenomena in the microwave region, such as quantum
squeezing \cite{Sqz_Movshovich90,Squeezing_Yurke05,SQZ_Buks05} and
experimental observation of the so called dynamical Casimir effect
\cite{segev06e}. These effects may also allow some intriguing technological
applications such as bifurcation amplifiers for quantum-limited measurements
\cite{BifAmp_Siddiqi04,BifAmp_Wiesenfeld86}, resonant readout of qubits
\cite{QBITS_Lee05}, mixers \cite{HEB_Floet99}, single photon detectors
\cite{HED_Goltsman05}, and more.

In this paper we study theoretically and experimentally the response of a
superconducting (SC) microwave stripline resonator, designed for enhanced
nonlinearity, to a monochromatic injected signal. We find that there is a
certain range of driving parameters, in which a novel nonlinear phenomenon
immerges, and self-sustained modulation (SM), of the reflected signal from the
resonator, is generated. That is, the resonator undergoes limited cycle
oscillations, ranging between several to tens of megahertz. Similar phenomenon
was briefly reported in the 60's
\cite{SM_Clorfeine64,DAiello66,SM_Peskovatskii67,SM_ERU70} in dielectric
resonators which were partially coated by a SC film, but it was not thoroughly
investigated and its significance was somewhat overlooked. This phenomenon is
of a significant importance as it introduces an extreme nonlinearity, which is
by far stronger than any other nonlinearity observed before in SC
resonators\cite{segev06d}. It results in a very strong intermodulation gain,
strong noise squeezing and period doubling of various orders\cite{segev06d},
strong coupling between different resonance modes\cite{segev06b}, and more.

The central results of this study have been recently reported in a short paper
\cite{segev06b}. In this paper we extend the previous report and derive a
theoretical model, according to which the SM originates by a thermal
instability in the resonator. The numerical integration of the model's
equations of motion exhibits SM, which have similar characteristics as the
experimental results. In addition, we derive analytic expressions for the
expected SM frequency and the spectral power density. These expressions also
yield a good agreement with the experimental results.

This paper is organized as follows. First we briefly describe the design of
our devices and the experimental setup. Then we present the SM phenomenon, as
measured in our devices. Afterwards, we derive the theoretical model, discuss
and justify its underlying assumptions and present the numerical integration
results. Finally, we quantitatively compare the predictions of the model to
typical experimental results.

\section{\label{CircuitDesign}Experimental Setup and Circuit Design}%

%TCIMACRO{\TeXButton{B}{\begin{figure*}
%\centering}}%
%BeginExpansion
\begin{figure*}
\centering
%EndExpansion%
\[
\text{%
%TCIMACRO{\FRAME{itbpF}{11.6361cm}{4.1163cm}{0cm}{}{}{fig1.eps}%
%maintain-aspect-ratio TRUE;  display "ICON";  valid_file "F";
%width 11.6361cm;  height 4.1163cm;  depth 0cm;  original-width 17.4727in;
%original-height 5.0261in;  cropleft "0";  croptop "1";  cropright "1";
%cropbottom "0";  filename 'fig1.eps';file-properties "XNPEU";}}}%
%BeginExpansion
\raisebox{-0cm}{\includegraphics[
height=4.1163cm,
width=11.6361cm
]%
{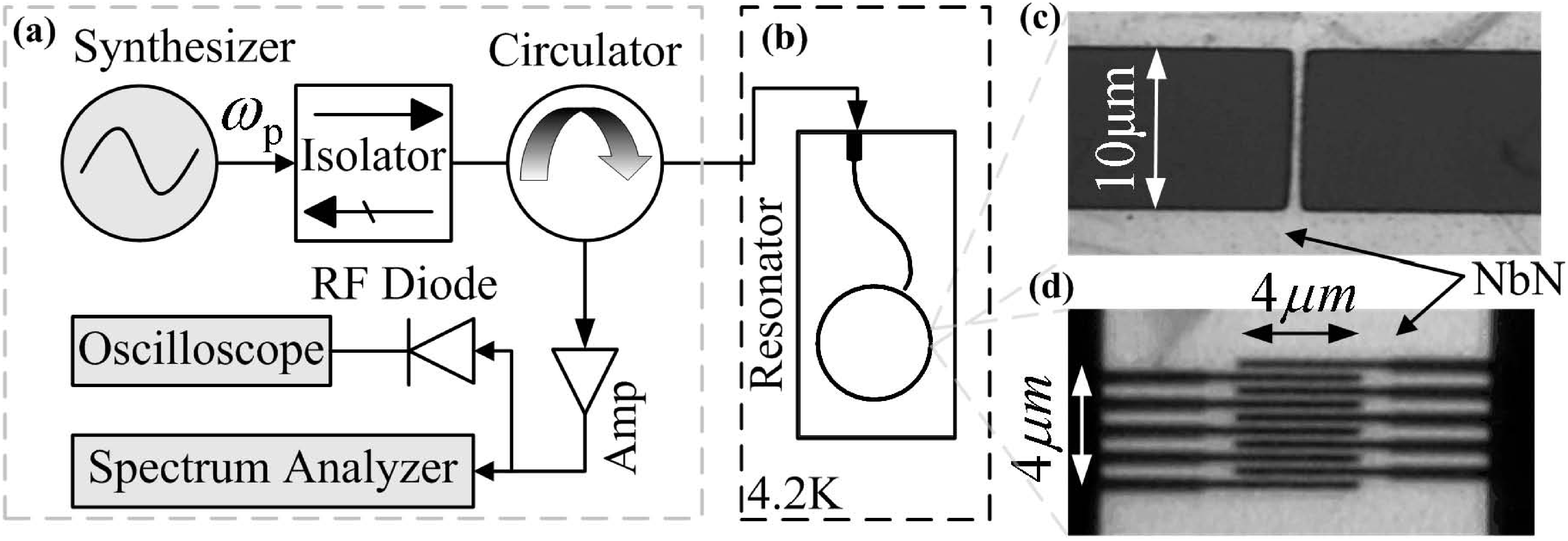}%
}%
%EndExpansion
}%
\]
%

%TCIMACRO{\TeXButton{caption}{\caption{$(\mathrm{a})$ SM measurement setup.
%$(\mathrm{b})$ Schematic layout of the device.
%Subplot $(\mathrm{c})$ and $(\mathrm{d}%
%)$ exhibit optical microscope images of the
%straight and meander shaped microbridges respectively.}}}%
%BeginExpansion
\caption{$(\mathrm{a})$ SM measurement setup.
$(\mathrm{b})$ Schematic layout of the device.
Subplot $(\mathrm{c})$ and $(\mathrm{d}%
)$ exhibit optical microscope images of the
straight and meander shaped microbridges respectively.}%
%EndExpansion
\label{ExpSetup_P_Device}%
%TCIMACRO{\TeXButton{E}{\end{figure*}} }%
%BeginExpansion
\end{figure*}
%EndExpansion
The majority of the experiments are performed using the experimental setup
described in Fig. \ref{ExpSetup_P_Device}$(\mathrm{a})$. The resonator is
stimulated with a monochromatic pump tone at an angular frequency
$\omega_{\mathrm{p}}$. The power reflected off the resonator is amplified at
room temperature and measured by using both a spectrum analyzer (SA) in the
frequency domain, and an oscilloscope, tracking the reflected power envelope
in the time domain. In other experiments, the $|S_{11}|$ reflection
coefficient is measured using a network analyzer (NA), connected directly to
the resonator RF port. All measurements are carried out while the device is
fully immersed in liquid Helium.
%TCIMACRO{\TeXButton{B}{\begin{figure*}
%\centering}}%
%BeginExpansion
\begin{figure*}
\centering
%EndExpansion%
\begin{tabular}
[c]{cc}%
%TCIMACRO{\FRAME{itbpFU}{2.8908in}{2.0872in}{0in}{\Qcb{$(\QTR{rm}{a})$}}%
%{}{fig2a.eps}{\special{ language "Scientific Word";  type "GRAPHIC";
%maintain-aspect-ratio TRUE;  display "ICON";  valid_file "F";
%width 2.8908in;  height 2.0872in;  depth 0in;  original-width 7.5741in;
%original-height 5.3939in;  cropleft "0";  croptop "1";  cropright "1";
%cropbottom "0";  filename 'fig2a.eps';file-properties "XNPEU";}}}%
%BeginExpansion
{\parbox[b]{2.8908in}{\begin{center}
\includegraphics[
height=2.0872in,
width=2.8908in
]%
{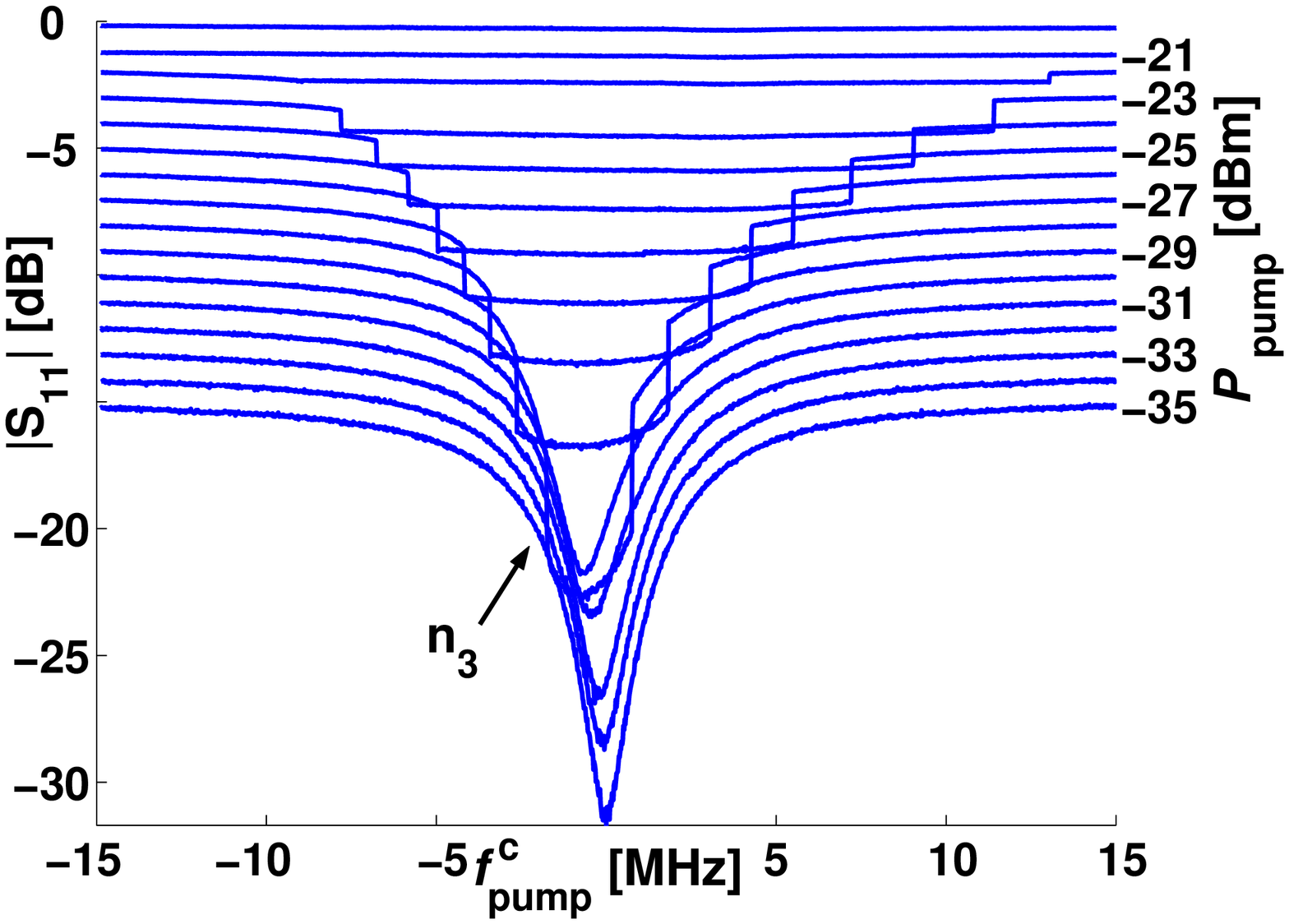}%
\\
$(\mathrm{a})$%
\end{center}}}%
%EndExpansion
&
%TCIMACRO{\FRAME{itbpFU}{2.8908in}{2.1179in}{0in}{\Qcb{$(\QTR{rm}{b})$}}%
%{}{fig2b.eps}{\special{ language "Scientific Word";  type "GRAPHIC";
%maintain-aspect-ratio TRUE;  display "ICON";  valid_file "F";
%width 2.8908in;  height 2.1179in;  depth 0in;  original-width 7.5741in;
%original-height 5.3649in;  cropleft "0";  croptop "1";  cropright "1";
%cropbottom "0";  filename 'fig2b.eps';file-properties "XNPEU";}}}%
%BeginExpansion
{\parbox[b]{2.8908in}{\begin{center}
\includegraphics[
height=2.1179in,
width=2.8908in
]%
{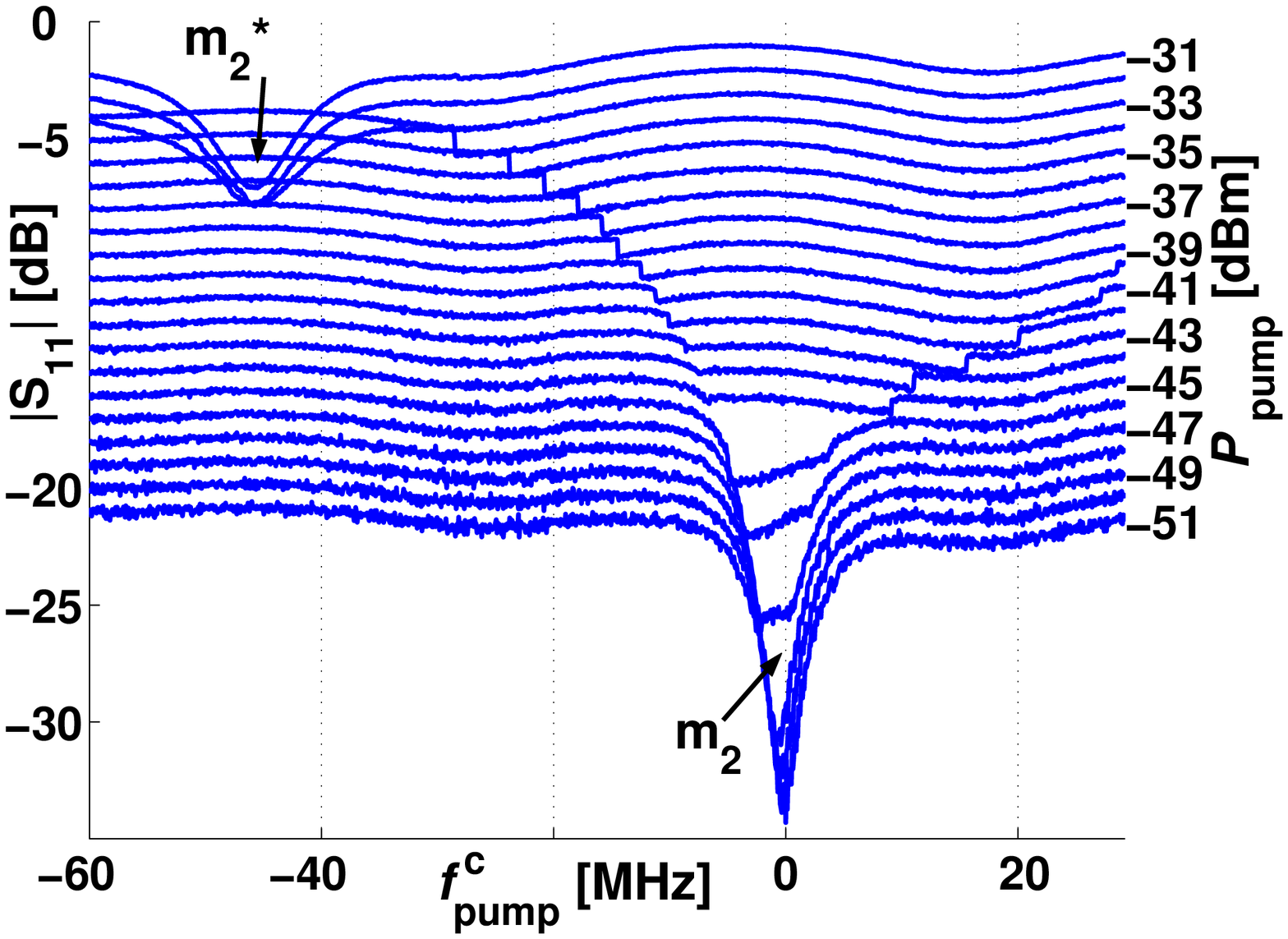}%
\\
$(\mathrm{b})$%
\end{center}}}%
%EndExpansion
\end{tabular}
%

%TCIMACRO{\TeXButton{caption}{\caption{(Color Online) $|S_{11}%
%|$ reflection measurements taken with $(\mathrm{a})$ E15
%and $(\mathrm{b})$ E16 devices. Note that the $|S_{11}|$ reflection
%coefficient is only defined for the case of a steady state reflection from a device.
%Therefore the $|S_{11}%
%|$ measurements taken at the SM region should be interpreted as an average  over
%time of the $|S_{11}|$ coefficient.}}}%
%BeginExpansion
\caption{(Color Online) $|S_{11}%
|$ reflection measurements taken with $(\mathrm{a})$ E15
and $(\mathrm{b})$ E16 devices. Note that the $|S_{11}|$ reflection
coefficient is only defined for the case of a steady state reflection from a device.
Therefore the $|S_{11}%
|$ measurements taken at the SM region should be interpreted as an average  over
time of the $|S_{11}|$ coefficient.}%
%EndExpansion
\label{S11_CollapsResFreq}%
%TCIMACRO{\TeXButton{E}{\end{figure*}}}%
%BeginExpansion
\end{figure*}%
%EndExpansion

A simplified circuit layout of the device is illustrated in Fig.
\ref{ExpSetup_P_Device}$(\mathrm{b})$. The resonator is designed as a
stripline ring \cite{Segev06a,NormRR_Chang87}, having a characteristic
impedance of $50%
%TCIMACRO{\unit{\U{3a9}}}%
%BeginExpansion
\operatorname{\Omega }%
%EndExpansion
$. It is composed of Niobium Nitride (NbN) deposited on a Sapphire wafer. The
first few resonance frequencies fall within the range of $2-8%
%TCIMACRO{\unit{GHz}}%
%BeginExpansion
\operatorname{GHz}%
%EndExpansion
$. A feedline, weakly coupled to the resonator, is employed for delivering the
input and output signals. A microbridge, which is employed as a weak-link that
allows the manipulation of the resonator's resonance frequencies
\cite{supRes_Saeedkia05}, is monolithically integrated into the structure of
the ring. Its angular location, relative to the feedline coupling location,
maximizes the RF current amplitude flowing through it in one of the resonance
modes, and thus maximizes its coupling to that mode. Further design
considerations, fabrication details as well as normal modes calculation can be
found elsewhere \cite{Segev06a}.

The results presented herein are obtained using two distinct devices, labeled
as E15 and E16, which differ by the geometry of the microbridges and by their
thicknesses. E15 has a $1\times10%
%TCIMACRO{\unit{\U{3bc}m}}%
%BeginExpansion
\operatorname{\mu m}%
%EndExpansion
^{2}$ microbridge geometry (Fig. \ref{ExpSetup_P_Device}$(\mathrm{c})$) and a
thickness of $200%
%TCIMACRO{\unit{nm}}%
%BeginExpansion
\operatorname{nm}%
%EndExpansion
$, whereas E16 has a $4\times4%
%TCIMACRO{\unit{\U{3bc}m}}%
%BeginExpansion
\operatorname{\mu m}%
%EndExpansion
^{2}$ meander shaped microbridge geometry (Fig. \ref{ExpSetup_P_Device}%
$(\mathrm{d})$) and a thickness of $8%
%TCIMACRO{\unit{nm}}%
%BeginExpansion
\operatorname{nm}%
%EndExpansion
$. The meander consists of nine strips, where each strip has a characteristic
area of $0.15\times4%
%TCIMACRO{\unit{\U{3bc}m}}%
%BeginExpansion
\operatorname{\mu m}%
%EndExpansion
^{2}$ and the strips are separated one from another by approximately $0.25%
%TCIMACRO{\unit{\U{3bc}m}}%
%BeginExpansion
\operatorname{\mu m}%
%EndExpansion
$ \cite{HED_Zhang03}.

The difference in the frequency response between E15 and E16 can be observed
in a simple $|S_{11}|$ reflection measurement, obtained using a NA. Fig.
\ref{S11_CollapsResFreq} shows various $|S_{11}|$ curves as a function of the
pump frequency centralized on the third resonance frequency $f_{3}=5.666%
%TCIMACRO{\unit{GHz}}%
%BeginExpansion
\operatorname{GHz}%
%EndExpansion
$ of E15 (panel $(\mathrm{a})$) \ and the second resonance frequency
$f_{2}=3.87%
%TCIMACRO{\unit{GHz}}%
%BeginExpansion
\operatorname{GHz}%
%EndExpansion
$ of E16 (panel $(\mathrm{b})$), labeled by $n_{3}$ and $m_{2}$ respectively.
Each curve represents a measurement with a different pump input power. For
clarity, the curves are vertically shifted upwards, for increasing power
values. The anomaly of the response is described as follows. Above some power
threshold the $|S_{11}|$ line-shapes cease having a normal Lorenzian shape,
their values substantially increase, and the resonance curves substantially
broaden and have steep edges. This behavior continues and intensifies as the
pump power further increases, until eventually no resonance is detected.
Furthermore, as seen in panel $(\mathrm{b})$, at relatively high power levels,
the resonance curve of E16 is reconstructed at a new resonance frequency, red
shifted by approximately $45%
%TCIMACRO{\unit{MHz}}%
%BeginExpansion
\operatorname{MHz}%
%EndExpansion
$ relative to $f_{2}$. The Lorenzian line-shape of the new resonance
frequency, labeled by $m_{2}^{\ast}$, represents a linear behavior in that
power range. These experimental results suggest that external stimulation can
cause a significant resonance shift in E16, while E15 can only experience an
increase in its damping rate.

\section{SM\ Experimental Observation}%

%TCIMACRO{\TeXButton{B}{\begin{figure*}
%\centering}}%
%BeginExpansion
\begin{figure*}
\centering
%EndExpansion%
%TCIMACRO{\FRAME{itbpF}{6.3163in}{2.6393in}{0in}{}{}{fig3.eps}%
%maintain-aspect-ratio TRUE;  display "ICON";  valid_file "F";
%width 6.3163in;  height 2.6393in;  depth 0in;  original-width 10.3362in;
%original-height 4.2922in;  cropleft "0";  croptop "1";  cropright "1";
%cropbottom "0";  filename 'fig3.eps';file-properties "XNPEU";}}}%
%BeginExpansion
{\includegraphics[
height=2.6393in,
width=6.3163in
]%
{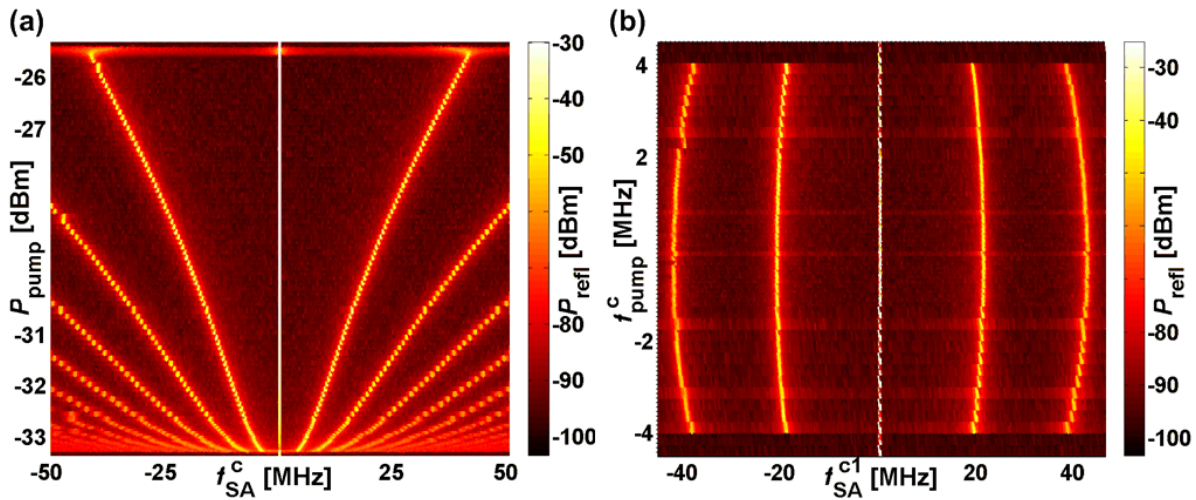}%
}%
%EndExpansion%
%TCIMACRO{\TeXButton{caption}{\caption{(Color online)
%Typical experimental results of the SM phenomenon in the frequency domain.
%Panel $(\mathrm{a})$ plots a colormap of the reflected power $P_{\mathrm
%{refl}}$
%as a function of the pump power $P_{\mathrm{pump}}$
%and the measured frequency $f_{\mathrm{SA}}$
%centralized on the third resonance frequency $f_3$ $(f_{\mathrm{SA}}%
%^{\mathrm{c}}=f_{\mathrm{SA}}-f_3)$ of E15,
%while the resonator is stimulated with a monochromatic pump at $f_3$.
%Panel $(\mathrm{b})$ plots a colormap of the
%reflected power as a function of the centralized measured frequency around the
%pump frequency $f_{\mathrm{SA}}^{\mathrm{c1}}=f_{\mathrm{SA}}-\omega
%_{\mathrm{p}}/2\protect\pi$
%and the centralized pump frequency around the resonance frequency
%$f_{\mathrm{pump}}^{\mathrm{c}}=\omega_{\mathrm{p}}/2\protect\pi
%-f_{3}$. The resonator
%is stimulated by a monochromatic pump having a power of $P_{\mathrm{pump}%
%}=-29.35~$dBm
%which drives the resonator into the regular SM zone.
%}}}%
%BeginExpansion
\caption{(Color online)
Typical experimental results of the SM phenomenon in the frequency domain.
Panel $(\mathrm{a})$ plots a colormap of the reflected power $P_{\mathrm
{refl}}$
as a function of the pump power $P_{\mathrm{pump}}$
and the measured frequency $f_{\mathrm{SA}}$
centralized on the third resonance frequency $f_3$ $(f_{\mathrm{SA}}%
^{\mathrm{c}}=f_{\mathrm{SA}}-f_3)$ of E15,
while the resonator is stimulated with a monochromatic pump at $f_3$.
Panel $(\mathrm{b})$ plots a colormap of the
reflected power as a function of the centralized measured frequency around the
pump frequency $f_{\mathrm{SA}}^{\mathrm{c1}}=f_{\mathrm{SA}}-\omega
_{\mathrm{p}}/2\protect\pi$
and the centralized pump frequency around the resonance frequency
$f_{\mathrm{pump}}^{\mathrm{c}}=\omega_{\mathrm{p}}/2\protect\pi
-f_{3}$. The resonator
is stimulated by a monochromatic pump having a power of $P_{\mathrm{pump}%
}=-29.35~$dBm
which drives the resonator into the regular SM zone.
}%
%EndExpansion
\label{e15_sm_pref3d}%
%TCIMACRO{\TeXButton{E}{\end{figure*}}}%
%BeginExpansion
\end{figure*}%
%EndExpansion

We now turn to investigate the region where SM immerges. Fig.
\ref{e15_sm_pref3d} shows typical experimental results of the SM phenomenon in
the frequency domain, as measured with E15. The dependence of the SM on the
pump power is shown in panel $(\mathrm{a})$ and described as follows. At low
input powers, approximately below $-33.25~$dBm, and at high input powers,
approximately above $-25.0~$dBm, the response of the resonator is linear,
namely, the reflected power from the resonator contains a single spectral
component at the frequency of the stimulating pump tone $\omega_{\mathrm{p}}$.
In between the two linear regions, there is a rather large power range in
which regular SM of the reflected power from the resonator occurs. It is
realized by rather strong and sharp sidebands, which extend over several
hundreds megahertz at both sides of the resonance frequency. The SM frequency,
which is defined as the frequency difference between the pump and the primary
sideband, increases as the pump power increases.

The regular SM starts and ends at two power thresholds, referred to as the
first and the second power thresholds. The first power threshold occurs at a
very narrow power range of approximately $10%
%TCIMACRO{\unit{nW}}%
%BeginExpansion
\operatorname{nW}%
%EndExpansion
$, during which the resonator's response desists being linear. It experiences
a strong amplification of the noise floor, known also as noise rise, over a
rather large frequency band, especially around the resonance frequency itself.
The second power threshold occurs on a slightly larger power range than the
first one and has similar, but less extreme characteristics.

As shown in panel $(\mathrm{b})$, the dependence of the SM on the pump
frequency is rather symmetric around the resonance frequency. It occurs only
within a well defined frequency range around the resonance frequency. A small
change in the pump frequency can abruptly ignite or quench the SM. Once
started though, the modulation frequency has a relatively weak dependence on
the pump frequency. E16 exhibits similar behavior, with slightly different
properties, as discussed below in section \ref{Stability}.

\section{Thermal Instability}

In this section we propose a theoretical model according to which the SM
originates by a thermal instability in the SC stripline resonator.
Current-carrying superconductors are known to have two metastable phases
sustained by Joule self-heating \cite{SM_Self-heatingNormalMetalsSuperconduct}%
. One phase is the SC phase and the other is an electrothermal local phase,
known as hotspot, which is basically an island of normal-conducting (NC)
domain, with a temperature above the critical one, surrounded by a SC domain.
This phenomenon can be explained by the heat balance equation holding at more
than one temperature. Due to an external \cite{HED_Kadin96} or internal
\cite{HED_Kitaygorsky05} perturbation, the hotspot can recover to the SC phase
or vice versa and thus oscillates between these phases. Such self-sustained
oscillations were often observed in experiments, for the case of a SC
microbridge driven by an external dc voltage or current (see review
\cite{SM_Self-heatingNormalMetalsSuperconduct} and references therein).

In the present case, as the microbridge is integrated into a stripline
resonator, the system is driven into instability via externally injected
microwave pump tone. Nonlinearity, according to our simple theoretical model,
results from the coupling between the equation of motion of the mode amplitude
in the resonator (Eq. (\ref{dB/dt})), and the thermal balance equation (Eq.
(\ref{dT/dt})) in the microbridge. The mechanism presented in this model is
somewhat similar to one of the mechanisms which cause self oscillations in an
optical parametric oscillator \cite{OPO_Suret00}.\ 

\subsection{Equations of Motion}

This section presents the equations of motion of the mode amplitude in the
resonator and the thermal balance in the microbridge. Note that Eqs.
(\ref{dB/dt})-(\ref{dT/dt}), (\ref{db/d_tau})-(\ref{Theta_inf}) also appear in
Ref. \cite{Baleegh06a}, and are re-presented here to ensure that the paper is self-contained.

\subsubsection{Mode Amplitude}%

%TCIMACRO{\FRAME{ftbpFU}{3.0353in}{1.5459in}{0pt}{\Qcb{(Color online) Schematic
%model of the driven resonator.}}{\Qlb{resonatormodeldiag}}{fig4.eps}%
%maintain-aspect-ratio TRUE;  display "ICON";  valid_file "F";
%width 3.0353in;  height 1.5459in;  depth 0pt;  original-width 11.3349in;
%original-height 5.7717in;  cropleft "0";  croptop "1";  cropright "1";
%cropbottom "0";  filename 'fig4.eps';file-properties "XNPEU";}}}%
%BeginExpansion
\begin{figure}
[ptb]
\begin{center}
\includegraphics[
height=1.5459in,
width=3.0353in
]%
{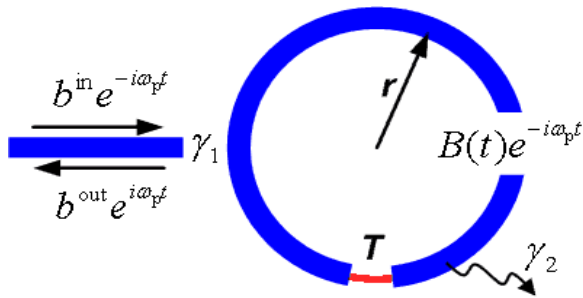}%
\caption{(Color online) Schematic model of the driven resonator.}%
\label{resonatormodeldiag}%
\end{center}
\end{figure}
%EndExpansion

Consider a resonator that is driven by a weakly coupled feedline carrying an
incident coherent tone $a^{\mathrm{in}}=b^{\mathrm{in}}e^{-i\omega
_{\mathrm{p}}t}$, where $b^{\mathrm{in}}$ is a constant complex amplitude
($\left\vert b^{\mathrm{in}}\right\vert ^{2}\propto P_{\mathrm{pump}}$, where
$P_{\mathrm{pump}}$ is the driving power) and $\omega_{\mathrm{p}}$ is the
driving angular frequency (See Fig. \ref{resonatormodeldiag}). The mode
amplitude inside the resonator can be written as $A=B\left(  t\right)
e^{-i\omega_{\mathrm{p}}t}$, where $B\left(  t\right)  $ is a complex
amplitude which is assumed to vary slowly on a time scale of $1/\omega
_{\mathrm{p}}$. In this approximation, the equation of motion of $B$ reads
\cite{Squeezing_Yurke05}%

\begin{equation}
\frac{\mathrm{d}B}{\mathrm{d}t}=\left[  i\left(  \omega_{\mathrm{p}}%
-\omega_{0}\left(  T\right)  \right)  -\gamma\left(  T\right)  \right]
B-i\sqrt{2\gamma_{1}}b^{\mathrm{in}}+c^{\mathrm{in}}, \label{dB/dt}%
\end{equation}
where $\omega_{0}\left(  T\right)  $ is the temperature dependant angular
resonance frequency, $T$ is the temperature of the hotspot, $\gamma\left(
T\right)  =\gamma_{1}+\gamma_{2}\left(  T\right)  $, where $\gamma_{1}$ is the
coupling constant between the resonator and the feedline, and $\gamma
_{2}\left(  T\right)  $ is the temperature dependant damping rate of the mode.

The term $c^{\mathrm{in}}$ represents an input Gaussian noise with a zero-mean
and a random phase, thus $\langle c^{\mathrm{in}}\rangle=0$, $\langle
c^{\mathrm{in}}(t)c^{\mathrm{in}}(t^{\prime})\rangle=\langle c^{in\ast
}(t)c^{in\ast}(t^{\prime})\rangle=0$, and its autocorrelation function is
given by $\langle c^{\mathrm{in}}(t)c^{in\ast}(t^{\prime})\rangle=G\omega
_{0}\delta\left(  t-t^{\prime}\right)  $. Consider the case of relatively high
temperature, $k_{B}T_{\mathrm{eff}}\gg\hbar\omega_{0}$, where $k_{B}$ is
Boltzmann's constant, and $T_{\mathrm{eff}}$ is a weighted average between $T$
and $T_{0}$, where $T_{0}$ is the temperature of the coolant and the weight
factors are discussed in Ref. \cite{segev06e}. Then at steady state, the
variance of the noise is given by $G=2\gamma k_{B}T_{\mathrm{eff}}/\hbar
\omega_{0}^{2}$.

The steady state solution of Eq. (\ref{dB/dt}), which is denoted as
$B_{\infty}$, is given by%

\begin{equation}
B_{\infty}=\frac{i\sqrt{2\gamma_{1}}b^{\mathrm{in}}}{i\left(  \omega
_{\mathrm{p}}-\omega_{0}\right)  -\gamma}. \label{Binf}%
\end{equation}

\subsubsection{Thermal Balance}

Consider the case where the nonlinearity originates by a local hotspot in the
resonator's microbridge. \ If the hotspot is assumed to be sufficiently small,
its temperature $T$ can be considered as homogeneous. \ The temperature of
other parts of the resonator is assumed be equal to that of the coolant
$T_{0}$. \ The power $Q$ heating up the hotspot is given by $Q=\kappa
Q_{\mathrm{t}}$, where $Q_{t}=\hslash\omega_{0}2\gamma_{2}\left\vert
B\right\vert ^{2}$ is the total power dissipated in the resonator, and
$0\leqslant\kappa\leqslant1$ represents the portion of the dissipated power
that is being absorbed by the microbridge. The heat balance equation reads%
\begin{equation}
C\frac{\mathrm{d}T}{\mathrm{d}t}=Q\left(  B\right)  -W, \label{dT/dt}%
\end{equation}
where $C$ is the thermal heat capacity, $W=H\left(  T-T_{0}\right)  $ is the
heat transfer power to the coolant, and $H$ is the heat transfer coefficient. \ 

The steady state solution of Eq. (\ref{dT/dt}), which is denoted as
$T_{\infty}$, for $B=B_{\infty}$, is given by%

\begin{equation}
T_{\infty}=T_{0}+\frac{\kappa\hslash\omega_{0}2\gamma_{2}\left\vert B_{\infty
}\right\vert ^{2}}{H}%
\end{equation}

\subsubsection{Stability zones}

\paragraph{Coupling mechanism}

The coupling mechanism between Eq. (\ref{dB/dt}) and Eq. (\ref{dT/dt}) is
based on the dependence of both the resonance frequency and the damping rate
of the driven mode on the resistance of the microbridge
\cite{supRes_Saeedkia05}, which in turn depends on the temperature
$T$\cite{Segev06a}.

Here we assume the simplest case, where the resonance frequency $\omega_{0}$,
and the damping rate $\gamma_{2}$ have a step function dependence on the
temperature of the hotspot (the step occurs at the critical temperature
$T_{\mathrm{c}}$). It is based on the fact that recent experiments with
photodetectors, based on a thin layer of NbN, have demonstrated an intrinsic
switching time on the order of $30%
%TCIMACRO{\unit{ps}}%
%BeginExpansion
\operatorname{ps}%
%EndExpansion
$ (see Ref. \cite{HED_Goltsman05} and references therein). In addition, when
illuminating our devices with a power modulated infrared light in a similar
way to the experiment described in Ref. \cite{Segev06a}, we measure a clear
response up to modulation frequencies of several gigahertz with
E16\cite{segev06e} and several hundred megahertz with E15. Thus the transition
through the instability point is very fast on the time scale of the SM
frequency. We further assume that all other parameters in the model are
temperature independent.

Under this assumption, the heat generation on the temperature $T$ have
step-like dependence\cite{SM_Self-heatingNormalMetalsSuperconduct}. The system
thus may have in general up to two locally stable steady states, corresponding
to the SC and NC phases. \ A SC steady state exists when $T_{\infty}<T_{c}$,
or alternatively when $\left\vert B_{\infty}\right\vert ^{2}<E_{\mathrm{s}}$,
where $E_{\mathrm{s}}=H\left(  T_{\mathrm{c}}-T_{0}\right)  /2\kappa
\gamma_{2\mathrm{s}}\hslash\omega_{0}$. \ Similarly, a NC steady state exists
when $T_{\infty}>T_{c}$, or alternatively when $\left\vert B_{\infty
}\right\vert ^{2}>E_{\mathrm{n}}$, where $E_{\mathrm{n}}=H\left(
T_{\mathrm{c}}-T_{0}\right)  /2\kappa\gamma_{2\mathrm{n}}\hslash\omega_{0}$,
where the subscript $(\mathrm{s})$ and $(\mathrm{n})$ denote the value of the
corresponding parameter when the system is in the SC and the NC phases respectively.

\paragraph{\label{Stability}Stability}%

%TCIMACRO{\FRAME{ftbpFU}{3.4006in}{3.7426in}{0pt}{\Qcb{Stability zones of the
%system, corresponding to $(\QTR{rm}{a})$ E15 and $(\QTR{rm}{b})$ E16 related
%assumptions. The solid lines are obtained from Eq. (\ref{Binf}) using
%$\left\vert B_{\infty}\right\vert ^{2}=E_{\QTR{rm}{s}}$ and $\left\vert
%B_{\infty}\right\vert ^{2}=E_{\QTR{rm}{n}}$.}}{\Qlb{StabilityZone}}%
%{fig5.eps}{\special{ language "Scientific Word";  type "GRAPHIC";
%maintain-aspect-ratio TRUE;  display "ICON";  valid_file "F";
%width 3.4006in;  height 3.7426in;  depth 0pt;  original-width 13.6247in;
%original-height 15.007in;  cropleft "0";  croptop "1";  cropright "1";
%cropbottom "0";  filename 'fig5.eps';file-properties "XNPEU";}}}%
%BeginExpansion
\begin{figure}
[ptb]
\begin{center}
\includegraphics[
height=3.7426in,
width=3.4006in
]%
{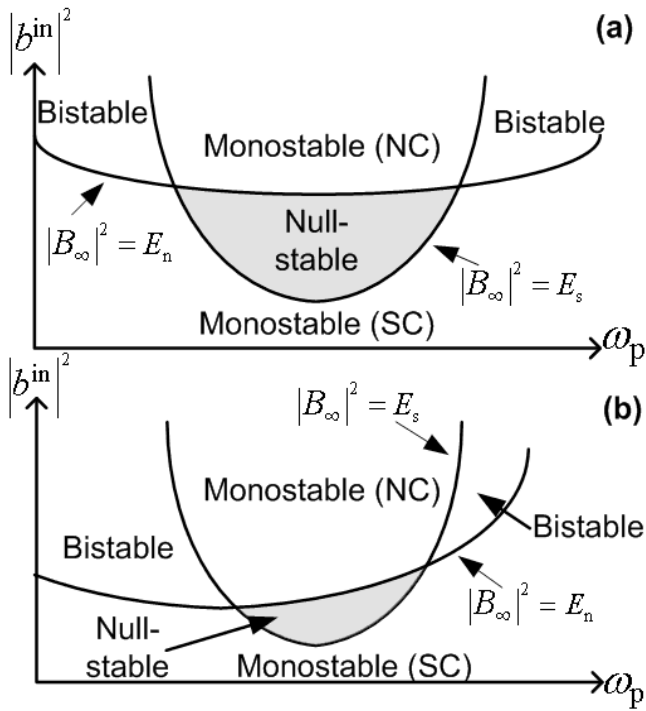}%
\caption{Stability zones of the system, corresponding to $(\mathrm{a})$ E15
and $(\mathrm{b})$ E16 related assumptions. The solid lines are obtained from
Eq. (\ref{Binf}) using $\left\vert B_{\infty}\right\vert ^{2}=E_{\mathrm{s}}$
and $\left\vert B_{\infty}\right\vert ^{2}=E_{\mathrm{n}}$.}%
\label{StabilityZone}%
\end{center}
\end{figure}
%EndExpansion
The stability of each of these phases depends on both the power and frequency
parameters of the injected pump tone, as described by the stability diagrams
in Fig. \ref{StabilityZone}. Panel $(\mathrm{a})$ shows a case similar to E15,
where the SC and the NC phases differ by the damping rate value, which has a
larger value in the latter phase. Panel $(\mathrm{b})$, on the other hand,
shows a case similar to E16, where the resonance frequency in the SC phase is
higher than the one in the NC phase, whereas the damping rates are the same in
both phases.

Four different stability zones can be identified in the diagram. In the
monostable zone either the SC phase or the NC phase is locally stable, whereas
in the bistable zones both phases are locally
stable\cite{Baleegh_bifurcation,Baleegh06a}. In the nullstable zone, on the
other hand, none of the phases are locally stable, and the resonator is
expected to oscillate between these two phases. As the two phases
significantly differ in their reflection coefficients, the oscillations are
translated into a modulation of the reflected pump tone. Note that, the
stability diagram indicates the existence of both power and frequency
hysteresis in the system's response\cite{Baleegh06a,Baleegh_bifurcation}.
Furthermore, panel $(\mathrm{b})$ in fig. \ref{StabilityZone} shows that the
dependence of the SM on the pump frequency is asymmetric in the case where a
resonance frequency shift occurs.

\begin{figure*}
\centering
%EndExpansion%
%TCIMACRO{\FRAME{itbpF}{6.3163in}{2.4051in}{0in}{}{}{fig6.eps}%
%maintain-aspect-ratio TRUE;  display "ICON";  valid_file "F";
%width 6.3163in;  height 2.4051in;  depth 0in;  original-width 10.4707in;
%original-height 3.9576in;  cropleft "0";  croptop "1";  cropright "1";
%cropbottom "0";  filename 'fig6.eps';file-properties "XNPEU";}}}%
%BeginExpansion
{\includegraphics[
height=2.4051in,
width=6.3163in
]%
{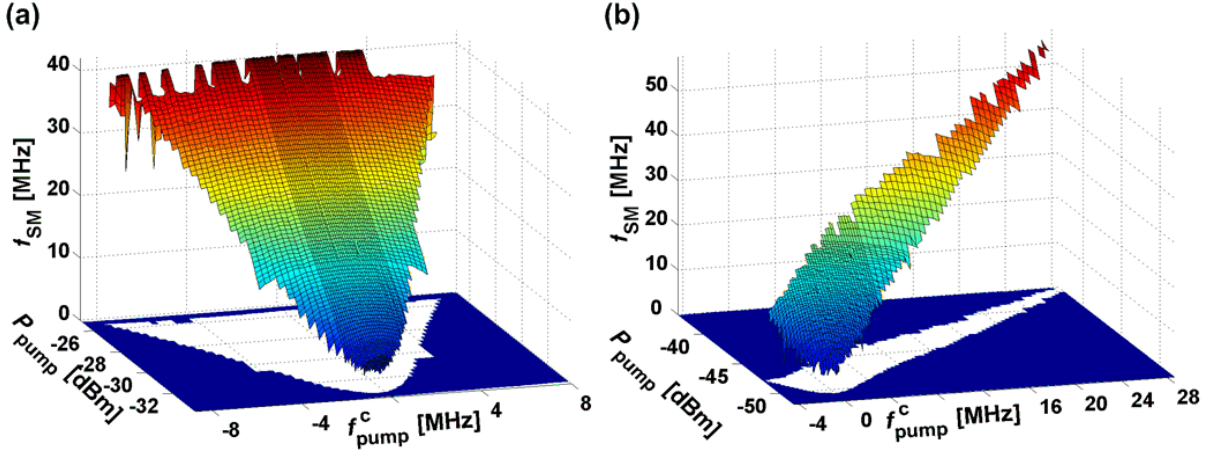}%
}%
%EndExpansion%
%TCIMACRO{\TeXButton{caption}{\caption
%{(Color online) Measured SM frequency $f_{\mathrm{sm}}%
%$ as a function of the pump
%power $P_{\mathrm{pump}}$
%and the pump frequency $f_{\mathrm{pump}}^{\mathrm{c}}%
%$,  centralized on $(\mathrm{a})$ the third
%and $(\mathrm{b}%
%)$ the second resonance frequencies of E15 and E16, respectively.}}}%
%BeginExpansion
\caption{(Color online) Measured SM frequency $f_{\mathrm{sm}}%
$ as a function of the pump
power $P_{\mathrm{pump}}$
and the pump frequency $f_{\mathrm{pump}}^{\mathrm{c}}%
$,  centralized on $(\mathrm{a})$ the third
and $(\mathrm{b}%
)$ the second resonance frequencies of E15 and E16, respectively.}%
%EndExpansion
\label{SM-Frequency}%
%TCIMACRO{\TeXButton{E}{\end{figure*}}}%
%BeginExpansion
\end{figure*}

This asymmetry is indeed observed experimentally when measuring the SM
frequency as a function of the pump frequency and power as shown in Fig.
\ref{SM-Frequency} panels $(\mathrm{a})$ and $(\mathrm{b})$, presenting data
obtained with E15 and E16 respectively. In both cases SM occurs in the
nullstability zone (compare with Fig. \ref{StabilityZone}). One clearly
notices that the SM, as measured with E16, is strongly asymmetric in frequency
in contrast to the case of E15. The maximum measured SM frequency is
approximately $41.1%
%TCIMACRO{\unit{MHz}}%
%BeginExpansion
\operatorname{MHz}%
%EndExpansion
$ and $57.6%
%TCIMACRO{\unit{MHz}}%
%BeginExpansion
\operatorname{MHz}%
%EndExpansion
$ with E15 and E16, respectively.
%TCIMACRO{\TeXButton{B}{\begin{figure*}
%\centering}}%
%BeginExpansion
%EndExpansion

\subsection{Adiabatic Approximation}

In the following section we derive two analytic expressions, one for the SM
period (Eq. (\ref{T})) and another for the SM spectral density (Eq.
(\ref{P_n neat onset})). Both expressions are valid for input powers that are
slightly larger than the first power threshold. The derivation assumes that
the system is in the adiabatic regime, namely that the rate in which the
temperature of the hotspot changes is much faster than the SM frequency. In
addition it assumes for simplicity that the pump frequency equals the
resonance frequency.

\subsubsection{Dimensionless variables}

In terms of the dimensionless time $\tau=\omega_{0}t$ Eq. (\ref{dB/dt}) reads%
\begin{equation}
\frac{\mathrm{d}b}{\mathrm{d}\tau}+\lambda b=\frac{c^{\mathrm{in}}}{\omega
_{0}}, \label{db/d_tau}%
\end{equation}
where $b=B-B_{\infty}$, and $\lambda=\left[  \gamma-i\left(  \omega
_{\mathrm{p}}-\omega_{0}\right)  \right]  /\omega_{0}$.

Defining the dimensionless temperature $\Theta=\left(  T-T_{0}\right)
/\left(  T_{\mathrm{c}}-T_{0}\right)  $, and using the dimensionless time
$\tau$ Eq. (\ref{dT/dt}) reads%
\begin{equation}
\frac{\mathrm{d}\Theta}{\mathrm{d}\tau}+g\left(  \Theta-\Theta_{\infty
}\right)  =0, \label{dT/d_tau}%
\end{equation}
where%

\begin{equation}
\Theta_{\infty}=\frac{2\hslash\kappa\gamma_{2}\left\vert B\right\vert ^{2}%
}{gC\left(  T_{\mathrm{c}}-T_{0}\right)  }=\frac{2\kappa\gamma_{2}%
\rho\left\vert B\right\vert ^{2}}{\omega_{0}g} \label{ThetaInf}%
\end{equation}
is the steady state value of the dimensionless temperature for a fixed mode
amplitude, and%

\begin{equation}
g=\frac{H}{C\omega_{0}}, \label{g_parameter}%
\end{equation}%
\begin{equation}
\rho=\frac{\hslash\omega_{0}}{C\left(  T_{\mathrm{c}}-T_{0}\right)  }.
\label{ro_parameter}%
\end{equation}
The steady states solution of Eq. (\ref{dT/d_tau}), where $B=B_{\infty}$, is
denoted as%
\begin{equation}
\Theta_{\infty0}=\frac{2\hslash\kappa\gamma_{2}\left\vert B_{\infty
}\right\vert ^{2}}{gC\left(  T_{c}-T_{0}\right)  }=\frac{2\kappa\gamma_{2}%
\rho\left\vert B_{\infty}\right\vert ^{2}}{\omega_{0}g}. \label{Theta_inf}%
\end{equation}

\subsubsection{Adiabatic Solution}

Assuming the case where $\Theta\neq1$ in the time interval $\left(
0,\tau\right)  $, and disregarding noise, the solution of Eq. (\ref{db/d_tau})
is given by%
\begin{equation}
B\left(  \tau\right)  =B_{\infty}\left[  1+\frac{B\left(  0\right)
-B_{\infty}}{B_{\infty}}\exp\left(  -\lambda\tau\right)  \right]  ,
\label{B(tau)}%
\end{equation}
thus, using the notation $\beta=\left(  B\left(  0\right)  -B_{\infty}\right)
/B_{\infty}$ and Eq. (\ref{Theta_inf}), one has%
\begin{align}
\Theta_{\infty}\left(  \tau^{\prime}\right)   &  =\Theta_{\infty0}[1+\beta
\exp\left(  -\lambda\tau^{\prime}\right)  +\nonumber\\
&  +\beta^{\ast}\exp\left(  -\lambda^{\ast}\tau^{\prime}\right)  +\nonumber\\
&  +\left\vert \beta\right\vert ^{2}\exp\left[  -\left(  \lambda+\lambda
^{\ast}\right)  \tau^{\prime}\right]  . \label{Tetha_inf1}%
\end{align}

In the adiabatic limit, where $\gamma/g\omega_{0}\ll1,$ one expects that the
temperature closely follows the evolution of the mode amplitude, namely
$\Theta\left(  \tau\right)  \simeq\Theta_{\infty}\left(  \tau\right)  $, thus
it is convenient to rewrite Eq. (\ref{dT/d_tau}) as%
\begin{equation}
\frac{\mathrm{d}\xi}{\mathrm{d}\tau}+g\xi=-\frac{\mathrm{d}\Theta_{\infty}%
}{\mathrm{d}\tau}, \label{d theta/ d tau}%
\end{equation}
where $\xi\left(  \tau\right)  =\Theta\left(  \tau\right)  -\Theta_{\infty
}\left(  \tau\right)  $. Using Eq. (\ref{Tetha_inf1}) the solution of Eq.
(\ref{d theta/ d tau}) for the case $g\tau\gg1$ can be written as
\begin{equation}
\Theta\left(  \tau\right)  =\Theta_{\infty}\left(  \tau\right)  -\frac{1}%
{g}\frac{\mathrm{d}\Theta_{\infty}\left(  \tau\right)  }{\mathrm{d}\tau}.
\label{ThetaAdiabatic}%
\end{equation}
Thus the lagging of the temperature $\Theta\left(  \tau\right)  $ behind the
asymptotic value $\Theta_{\infty}$ depends on the rate of change of
$\left\vert B\right\vert ^{2}$ (see Eq. (\ref{ThetaInf})).

Moreover, when the pump frequency equals the resonance frequency, namely
$\omega_{\mathrm{p}}=\omega_{0}$, $B$ is purely imaginary and its time
evolution is given by%
\begin{equation}
\frac{B\left(  t\right)  -B_{\infty}}{B\left(  0\right)  -B_{\infty}}%
=\exp\left(  -\frac{\gamma\tau}{\omega_{0}}\right)  , \label{B(t) res}%
\end{equation}
where $B_{\infty}=-i\sqrt{2\gamma_{1}}b^{\mathrm{in}}/\gamma$. Consequently
Eq. (\ref{ThetaAdiabatic}) reads%
\begin{equation}
\Theta\left(  \tau\right)  =\Theta_{\infty}\left(  \tau\right)  -\frac{\gamma
}{g\omega_{0}}\left[  \Theta_{\infty0}-\Theta_{\infty}\left(  \tau\right)
\right]  . \label{Theta(t)}%
\end{equation}

\subsubsection{Null-stability zone}

Consider the case of operating in the nullstability zone. Switching between SC
and NC phases occurs when $\Theta\left(  t\right)  =1$. \ At that time the
mode amplitude $B$ can be found from the value of $\Theta_{\infty}$. \ Using
Eq. (\ref{Theta(t)}) one finds to first order in $\gamma/g\omega_{0}$%
\begin{equation}
\Theta_{\infty}=1+\frac{\gamma\left(  \Theta_{\infty0}-1\right)  }{g\omega
_{0}}.
\end{equation}
Using Eq. (\ref{ThetaInf}) and Eq. (\ref{Theta_inf}) one finds that the mode
amplitude at switching, to first order in $\gamma/g\omega_{0}$, is given by%
\begin{equation}
\left\vert B\right\vert =\left\vert B_{0}\right\vert \left\{  1+\frac{\gamma
}{2g\omega_{0}}\left[  \left(  \frac{B_{\infty}}{B_{0}}\right)  ^{2}-1\right]
\right\}  . \label{BatSwitching}%
\end{equation}
where $\left\vert B_{0}\right\vert ^{2}=\omega_{0}g/2\kappa\gamma_{2}\rho$ is
the value of $\left\vert B\right\vert ^{2}$ for which $\Theta_{\infty}=1$. We
denote by $B_{\mathrm{s}}$ ($B_{\mathrm{n}}$) the value of $\left\vert
B\right\vert $ when a switching from the SC to the NC (NC to SC) phase occurs.

\paragraph{SM Period}

The SM period is denoted as $\mathcal{T=T}_{\mathrm{s}}\mathcal{+T}%
_{\mathrm{n}}$, where $\mathcal{T}_{\mathrm{s}}\mathcal{\ (T}_{\mathrm{n}}$)
is the time, in which the system is in the SC (NC) phase. \ Using Eq.
(\ref{B(t) res}) one finds%
\begin{equation}
\mathcal{T}_{\mathrm{s}}=\frac{1}{\gamma_{\mathrm{s}}}\log\frac{B_{\mathrm{n}%
}-B_{\infty\mathrm{s}}}{B_{\mathrm{s}}-B_{\infty\mathrm{s}}};~\mathcal{T}%
_{\mathrm{n}}=\frac{1}{\gamma_{\mathrm{n}}}\log\frac{B_{\mathrm{s}}%
-B_{\infty\mathrm{n}}}{B_{\mathrm{n}}-B_{\infty\mathrm{n}}}. \label{T_s}%
\end{equation}
Slightly above the first SM power threshold one has $B_{\mathrm{s}}\simeq
B_{\infty\mathrm{s}}$.\ In this case one expects that $\mathcal{T}%
_{\mathrm{s}}\gg\mathcal{T}_{\mathrm{n}}$. Moreover writing $\mathcal{T}%
_{\mathrm{s}}$ as%
\begin{equation}
\mathcal{T}_{\mathrm{s}}=\frac{1}{\gamma_{\mathrm{s}}}\left(  \log
\frac{B_{\infty\mathrm{s}}-B_{\mathrm{n}}}{B_{0\mathrm{s}}}+\log
\frac{B_{0\mathrm{s}}}{B_{\infty\mathrm{s}}-B_{\mathrm{s}}}\right)  ,
\end{equation}
where $B_{0\mathrm{s}}=-i\sqrt{2\gamma_{1\mathrm{s}}}b_{0}^{\mathrm{in}%
}/\gamma_{\mathrm{s}}$, $b_{0}^{\mathrm{in}}$ is the input amplitude
associated with the first SM power threshold, and neglecting the first term,
which is much smaller than the second one, and using Eq. (\ref{BatSwitching})
yield%
\begin{align}
\mathcal{T}  &  \simeq\mathcal{T}_{\mathrm{s}}\simeq\nonumber\\
&  -\frac{1}{\gamma_{\mathrm{s}}}\log\left\{  \frac{B_{\infty\mathrm{s}}%
}{B_{0\mathrm{s}}}-1-\frac{\gamma}{2g\omega_{0}}\left[  \left(  \frac
{B_{\infty}}{B_{0}}\right)  ^{2}-1\right]  \right\}  .
\end{align}
Thus, using the notation $\vartheta=\left(  b^{\mathrm{in}}-b_{0}%
^{\mathrm{in}}\right)  /b_{0}^{\mathrm{in}}$ one finds that slightly above the
first threshold, when $\vartheta\ll1$, namely when regular SM with a
relatively long period occurs, the SM period is given by%
\begin{equation}
\mathcal{T}\simeq\frac{1}{\gamma_{\mathrm{s}}}\log\frac{1}{\vartheta\left(
1-\frac{\gamma}{g\omega_{0}}\right)  }\simeq\frac{1}{\gamma_{\mathrm{s}}}%
\log\frac{1}{\vartheta}. \label{T}%
\end{equation}
Note that disregarding noise can not be justified very close to the first
power threshold, since in that region the system is extremely sensitive to fluctuations.

\paragraph{Spectral Density}

The output signal reflected from the resonator is written as $a^{\mathrm{out}%
}=b^{\mathrm{out}}e^{-i\omega_{\mathrm{p}}t}$, where $b^{\mathrm{out}}$ is a
complex amplitude. According to the input-output relation, which relates the
output signal to the input one\cite{Qant_InputOutputDampedQuantumSystems}, the
following holds%
\begin{equation}
\frac{b^{\mathrm{out}}}{\sqrt{\omega_{0}}}=\frac{b^{\mathrm{in}}}{\sqrt
{\omega_{0}}}-i\sqrt{\frac{2\gamma_{1}}{\omega_{0}}}B. \label{in out}%
\end{equation}

Above the first power threshold the amplitude $B\left(  t\right)  $ is
periodic, $B\left(  t\right)  =B\left(  t+\mathcal{T}\right)  $. If the
assumption $\mathcal{T}_{\mathrm{s}}\gg\mathcal{T}_{\mathrm{n}}$ holds, one
finds%
\begin{equation}
B\left(  t\right)  \simeq B_{\mathrm{n}}+\left(  B_{\infty\mathrm{s}%
}-B_{\mathrm{n}}\right)  \left(  1-e^{-\gamma_{\mathrm{s}}t}\right)  ,
\label{B(t) near onset}%
\end{equation}
where the time interval in which the hotspot is in NC phase is neglected. The
power spectrum of the $k^{\mathrm{th}}$ harmonic of $b^{\mathrm{out}}$ is
given by%
\begin{equation}
P_{k}=\frac{1}{\mathcal{T}}\left\vert
%TCIMACRO{\tint \limits_{0}^{\mathcal{T}}}%
%BeginExpansion
{\textstyle\int\limits_{0}^{\mathcal{T}}}
%EndExpansion
b^{\mathrm{out}}\left(  t\right)  e^{i\omega_{k}t}\ dt\right\vert ^{2},
\label{P_n}%
\end{equation}
where $\omega_{k}=2k\pi/\mathcal{T}$. Thus, using Eq. (\ref{in out}), the
spectral density slightly above the first SM power threshold is given by
\begin{equation}
P\left(  \omega_{k}\right)  =\frac{2\gamma_{1}\left(  B_{\mathrm{s}%
}-B_{\mathrm{n}}\right)  ^{2}}{\mathcal{T}\left(  \omega_{k}^{2}%
+\gamma_{\mathrm{s}}^{2}\right)  }. \label{P_n neat onset}%
\end{equation}

\subsubsection{Validity of the adiabatic approximation}

We now return to the adiabatic approximation $\gamma/g\omega_{0}\ll1$, and
examine its validity by estimating the value of the parameter $g$ in Eq.
(\ref{g_parameter}). Consider the case where the nonlinearity originates by a
hotspot of lateral area $A_{\mathrm{eff}}$, forming in the microbridge.\ The
heat capacity $C$ of the hotspot can be expressed as $C=C_{v}A_{\mathrm{eff}%
}d$, where $C_{v}$ is the heat capacity per unit volume, and $d$ is the
thickness of the NbN film. \ By further assuming that the generated heat is
cooled mainly down the substrate rather than along the
film\cite{kinInd_Johnson96}, the heat transfer coefficient reads $H=\alpha
A_{\mathrm{eff}}$, where $\alpha$ is the thermal surface conductance between
the NbN film and the substrate. According to this notation Eq.
(\ref{g_parameter}) is expressed as $g=\alpha/C_{v}d\omega_{0}$.\ To obtain an
estimate for the parameter $\rho$ in Eq. (\ref{ro_parameter}), we evaluate the
total dissipated power $Q_{\mathrm{t}}$ in the resonator at the first SM power
threshold $E_{\mathrm{s}}$, by assuming that the power dissipated in the
microbridge is given by $\kappa Q_{\mathrm{t}}=\kappa(1-\left\vert
S_{11}\right\vert _{\mathrm{th}})P_{\mathrm{pump}}$, where $P_{\mathrm{pump}}$
is the injected input power, $\left\vert S_{11}\right\vert _{\mathrm{th}}$ is
the reflection coefficient at the\ first SM\ power threshold and $\kappa
\simeq1$. On the other hand, this equals to the heat flow from the microbridge
to the substrate $W=\alpha A_{\mathrm{eff}}\left(  T-T_{0}\right)  $. Thus Eq.
(\ref{ro_parameter}) can be expressed as $\rho=\alpha\hslash\omega_{0}%
/C_{v}d\kappa Q_{\mathrm{t}}$. The value of $\alpha$ and $C_{v}$ parameters as
estimated for NbN on a Sapphire substrate at temperature $T=4.2%
%TCIMACRO{\unit{K}}%
%BeginExpansion
\operatorname{K}%
%EndExpansion
$ \cite{kinInd_Johnson96,HED_Weiser81}, are $\alpha\simeq12.5%
%TCIMACRO{\unit{W}}%
%BeginExpansion
\operatorname{W}%
%EndExpansion%
%TCIMACRO{\unit{cm}}%
%BeginExpansion
\operatorname{cm}%
%EndExpansion
^{-2}%
%TCIMACRO{\unit{K}}%
%BeginExpansion
\operatorname{K}%
%EndExpansion
^{-1}$ and $C_{v}\simeq2.7\times10^{-3}%
%TCIMACRO{\unit{J}}%
%BeginExpansion
\operatorname{J}%
%EndExpansion%
%TCIMACRO{\unit{cm}}%
%BeginExpansion
\operatorname{cm}%
%EndExpansion
^{-3}%
%TCIMACRO{\unit{K}}%
%BeginExpansion
\operatorname{K}%
%EndExpansion
^{-1}$.$\ $The various measured and calculated parameters are summarized in
table \ref{Model_Params}, which shows that the adiabatic assumption is
justified for E16, and is marginal for E15. But as the above estimation does
not take into account the direct contact between the sample and the liquid
Helium, it is reasonable to assume that the adiabatic assumption is justified
for E15 as well.\begin{table}[ptb]
\caption{Model's Parameters}%
\label{Model_Params}%
\renewcommand{\arraystretch}{1.5} \begin{ruledtabular}
\begin{tabular}{cccccc}
& E15 & E16 & & E15 & E16\\
\hline\hline
$\omega_{0}/2\protect\pi~[%
%TCIMACRO{\unit{GHz}}%
%BeginExpansion
\operatorname{GHz}%
%EndExpansion
]$ & $5.7$ & $3.8$ & \textit{Q}-factor & $880$ & $ 250$\\\hline
$d~[%
%TCIMACRO{\unit{nm}}%
%BeginExpansion
\operatorname{nm}%
%EndExpansion
]$ & $200$ & $8$ & $g~[\times10^{-3}]$ & $9.69$ & $250$\\\hline
$P_{\mathrm{pump}}~[$dBm$]$ & $-25.5$ & $-49$ &
$\rho~[\times10^{-9}]$ & $0.80$ & $1180$\\\hline
$\left\vert S_{11}\right\vert _{\mathrm{th}}~[$dB$]$ &
$-4.15$ & $-13$ & $\gamma/g\omega_{0}~[\times10^{-3}]$ & $197$ & $27$\\
\end{tabular}
\end{ruledtabular}\end{table}%

%TCIMACRO{\TeXButton{B}{\begin{figure*}
%\centering}}%
%BeginExpansion
\begin{figure*}
\centering
%EndExpansion%
%TCIMACRO{\FRAME{itbpF}{6.3146in}{1.6986in}{0in}{}{}{fig7.eps}%
%{\special{ language "Scientific Word";  type "GRAPHIC";  display "ICON";
%valid_file "F";  width 6.3146in;  height 1.6986in;  depth 0in;
%original-width 20.7928in;  original-height 5.5243in;  cropleft "0";
%filename 'fig7.eps';file-properties "XNPEU";}}}%
%BeginExpansion
{\includegraphics[
height=1.6986in,
width=6.3146in
]%
{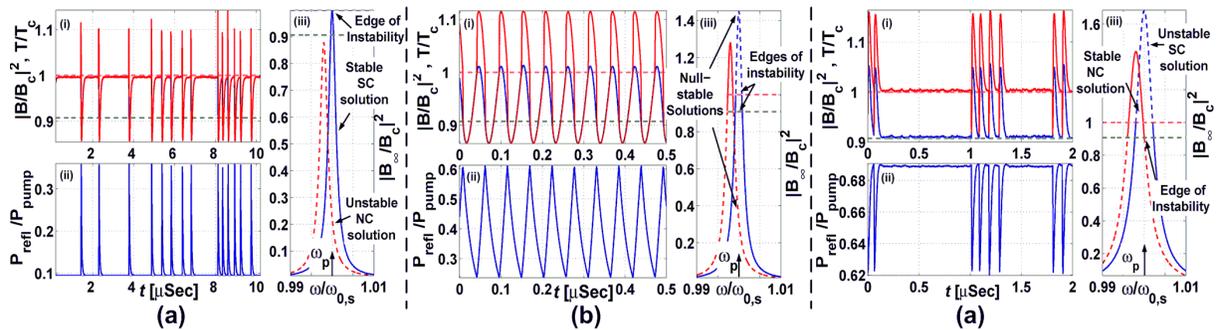}%
}%
%EndExpansion%
%TCIMACRO{\TeXButton{caption}{\caption
%{(Color online) Numerical integration of the equations of motion of the model at three distinct pump powers
%$(\mathrm{a})$ the first power threshold $(\mathrm{b}%
%)$ some intermediate power between the first and second
%power thresholds , and $(\mathrm{c}%
%)$ the second power threshold. Each panel has three subplots.
%Subplot (i) describes the normalized mode amplitude $B_{\mathrm{N}%
%}$ (blue), and the normalized hotspot
%temperature $T_{\mathrm{N}}$
%(red), as a function of time, where both are normalized by their critical values at which a transition from the SC to
%the NC phase occurs. Subplot (ii) plots the reflected power off the resonator $P_{\mathrm
%{refl}}$
%normalized by the impinging
%pump power $P_{\mathrm{pump}}%
%$. Subplot (iii) describes the normalized steady state solution of the mode
%amplitude (Eq.(\ref{Binf}%
%)), as a function of the normalized frequency, for the case where it is decoupled
%from the bridge temperature.
%The solution is again normalized by its critical value $E_{\mathrm{s}}$,
%and the frequency is normalized by the SC resonance
%frequency. The solid and dashed portions of the curves represent solutions which are stable and unstable
%respectively, according to the stability diagram in Fig. \ref{StabilityZone}%
%$(\mathrm{b})$.
%The blue and red curves are solutions for the cases where the system is in the SC and the NC phase, respectively.
%The magenta and green dashed lines in subplots (i,iii) show
%$E_{\mathrm{s}}$ and $E_{\mathrm{n}}$ normalized by $E_{\mathrm{s}%
%}$ respectively.}}}%
%BeginExpansion
\caption
{(Color online) Numerical integration of the equations of motion of the model at three distinct pump powers
$(\mathrm{a})$ the first power threshold $(\mathrm{b}%
)$ some intermediate power between the first and second
power thresholds , and $(\mathrm{c}%
)$ the second power threshold. Each panel has three subplots.
Subplot (i) describes the normalized mode amplitude $B_{\mathrm{N}%
}$ (blue), and the normalized hotspot
temperature $T_{\mathrm{N}}$
(red), as a function of time, where both are normalized by their critical values at which a transition from the SC to
the NC phase occurs. Subplot (ii) plots the reflected power off the resonator $P_{\mathrm
{refl}}$
normalized by the impinging
pump power $P_{\mathrm{pump}}%
$. Subplot (iii) describes the normalized steady state solution of the mode
amplitude (Eq.(\ref{Binf}%
)), as a function of the normalized frequency, for the case where it is decoupled
from the bridge temperature.
The solution is again normalized by its critical value $E_{\mathrm{s}}$,
and the frequency is normalized by the SC resonance
frequency. The solid and dashed portions of the curves represent solutions which are stable and unstable
respectively, according to the stability diagram in Fig. \ref{StabilityZone}%
$(\mathrm{b})$.
The blue and red curves are solutions for the cases where the system is in the SC and the NC phase, respectively.
The magenta and green dashed lines in subplots (i,iii) show
$E_{\mathrm{s}}$ and $E_{\mathrm{n}}$ normalized by $E_{\mathrm{s}%
}$ respectively.}%
%EndExpansion
\label{SM-Model}%
%TCIMACRO{\TeXButton{E}{\end{figure*}}}%
%BeginExpansion
\end{figure*}%
%EndExpansion

\section{Numerical Integration}

Following the discussion in section \ref{CircuitDesign}, the numerical results
presented for E15 are calculated by assuming a significant increase in the
damping rate and a negligible shift in the resonance frequency as the critical
temperature is exceeded, and the results presented for E16 are calculated by
assuming the opposite case, where only $\omega_{0}$ is temperature dependant.

Fig. \ref{SM-Model} shows numerical integration results of the coupled Eq.
(\ref{dB/dt}) and Eq. (\ref{dT/dt}), using the case where a shift in the
resonance frequency occurs (E16). Results obtained by assuming a significant
increase in the damping rate with no frequency shift are presented in Ref.
\cite{segev06b}. The resonator is stimulated by an impinging pump tone at the
SC resonance frequency. Panel $(\mathrm{a})$ shows results that are calculated
for a pump power in the first SM power threshold range. If the system would
have been noiseless then both the mode amplitude and the temperature could
reach a steady state. At this steady state the reflection from the resonator
is relatively low as the pump frequency coincides with the resonance
frequency. As this steady state is on the edge of instability, the thermal
noise at a temperature of $4.2%
%TCIMACRO{\unit{K}}%
%BeginExpansion
\operatorname{K}%
%EndExpansion
$ makes the system unstable and it occasionally falls off the edge and
switches to the NC phase. When this happens, the dissipation slightly
increases but more significantly, the resonance frequency shifts, and
consequently the mode amplitude starts decreasing. As a result the heat
production, which is proportional to the mode amplitude squared, decreases and
thus when the excess heat is transferred to the substrate, and the temperature
of the bridge decreases below the critical one, the resonator switches back to
the SC state and a new buildup cycle of electromagnetic energy begins.
Accordingly, the reflected power is low for most of the time, and these cycles
are realized as sporadic but correlated spikes of the reflected power. The
heat generated in such a spike, raises the probability for a sequential spike
to occur, and thus induces a positive correlation between the spikes
(bouncing). The dynamics of the relaxation cooldown of these spikes is
similar, and thus their line-shapes also. The power range in which these
spikes are triggered, and hence the width of the power threshold range, is
governed by the noise intensity.

Panel $(\mathrm{b})$ shows results calculated for some pump power in the range
of the regular SM. The evolution of the system is similar to the one just
described, with one major difference. When the system is in the SC phase the
mode amplitude is built toward a nullstable state and thus a steady state is
not achieved. As a result, regular oscillations occur without the assistance
of noise, which in general, has a negligible impact in that power range.

Panel $(\mathrm{c})$ shows results calculated for a pump power at the second
SM power threshold while assuming a $15%
%TCIMACRO{\unit{K}}%
%BeginExpansion
\operatorname{K}%
%EndExpansion
$ thermal noise and a slightly enhanced damping rate due to an average
increase in the microbridge temperature. The behavior of the system at this
threshold resembles the first one, but the SC and NC phases exchange roles. In
this power range the resonator is in the NC, high reflective phase for most of
the time and noise-induced spikes temporarily drive it to the SC, low
reflective phase. The internal thermal noise at the second threshold is
stronger than at the first one and consequently this power threshold range is wider.%

%TCIMACRO{\FRAME{ftbpFU}{3.4014in}{2.753in}{0pt}{\Qcb{(Color Online) SM
%reflected power line-shapes of experimental (blue) and the numerical
%integration of the model's equations of motion (dashed red) results,
%normalized by the maximum peak to peak value and the incident pump power
%respectively.}}{\Qlb{CompLineShape}}{fig8.eps}%
%maintain-aspect-ratio TRUE;  display "ICON";  valid_file "F";
%width 3.4014in;  height 2.753in;  depth 0pt;  original-width 6.7729in;
%original-height 5.4736in;  cropleft "0";  croptop "1";  cropright "1";
%cropbottom "0";  filename 'fig8.eps';file-properties "XNPEU";}}}%
%BeginExpansion
\begin{figure}
[ptb]
\begin{center}
\includegraphics[
height=2.753in,
width=3.4014in
]%
{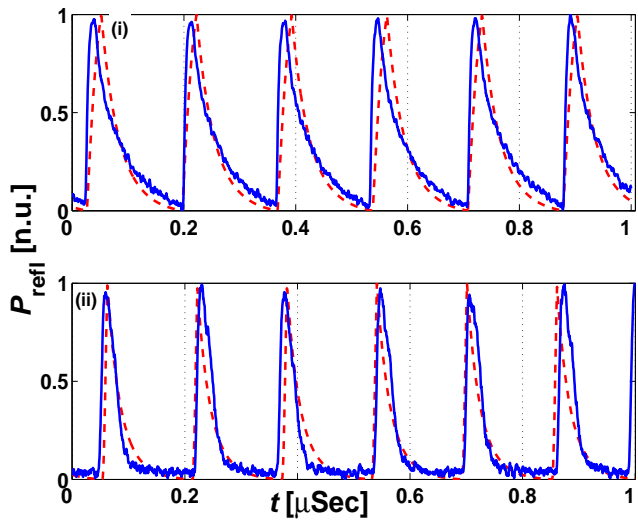}%
\caption{(Color Online) SM reflected power line-shapes of experimental (blue)
and the numerical integration of the model's equations of motion (dashed red)
results, normalized by the maximum peak to peak value and the incident pump
power respectively.}%
\label{CompLineShape}%
\end{center}
\end{figure}
%EndExpansion

Fig. \ref{CompLineShape} shows the envelope line-shape of the reflected power
when a regular SM, having a frequency of approximately $6%
%TCIMACRO{\unit{MHz}}%
%BeginExpansion
\operatorname{MHz}%
%EndExpansion
$, occurs. The experimental data is seen in blue, and the numerical
integration results are seen in dashed red. The two subplots include data
obtained with E15 (panel $(i)$) and E16 (panel $(ii)$), respectively. The
numerical results were calculated using the corresponding parameters for each
device as discussed above. The comparison shows a good match between the model
and the experimental data for both cases.

\section{Analytical Results}%

%TCIMACRO{\FRAME{ftbpFU}{3.4014in}{2.5662in}{0pt}{\Qcb{(Color Online) SM
%frequency as a function of the normalized injected power $\vartheta$, measured
%with E15 (upper) and E16 (lower) devices.}}{\Qlb{compareSMfreq2Model}%
%}{fig9.eps}{\special{ language "Scientific Word";  type "GRAPHIC";
%maintain-aspect-ratio TRUE;  display "ICON";  valid_file "F";
%width 3.4014in;  height 2.5662in;  depth 0pt;  original-width 6.8667in;
%original-height 5.1656in;  cropleft "0";  croptop "1";  cropright "1";
%cropbottom "0";  filename 'fig9.eps';file-properties "XNPEU";}}}%
%BeginExpansion
\begin{figure}
[ptb]
\begin{center}
\includegraphics[
height=2.5662in,
width=3.4014in
]%
{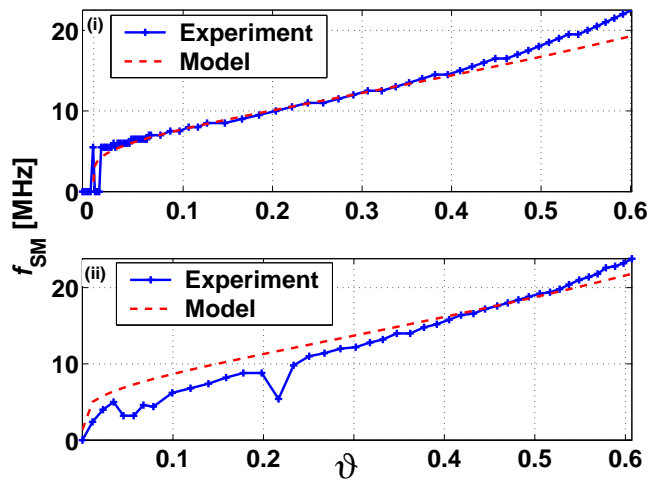}%
\caption{(Color Online) SM frequency as a function of the normalized injected
power $\vartheta$, measured with E15 (upper) and E16 (lower) devices.}%
\label{compareSMfreq2Model}%
\end{center}
\end{figure}
%EndExpansion

Fig. \ref{compareSMfreq2Model} shows a comparison between the measured SM
frequency and the one predicted by Eq. (\ref{T}), for data taken with E15
(panel $(i)$) and E16 (panel $(ii)$) devices. Eq. (\ref{T}) is expected to
hold when the input power is slightly above the first SM power threshold.
Indeed, when using an experimentally measured damping rate of $\gamma
_{\mathrm{s}}=26%
%TCIMACRO{\unit{MHz}}%
%BeginExpansion
\operatorname{MHz}%
%EndExpansion
$ for E16\cite{Segev06a} and a fitted damping rate of $\gamma_{\mathrm{s}%
}=23.1%
%TCIMACRO{\unit{MHz}}%
%BeginExpansion
\operatorname{MHz}%
%EndExpansion
$ for E15, the model yields a good agreement for both devices.%
%TCIMACRO{\FRAME{ftbpFU}{3.4006in}{2.4574in}{0pt}{\Qcb{(Color Online) The solid
%blue curves in the upper and lower subplots show typical SM experimental
%results in the frequency domain, obtained with E15 and E16 respectively. The
%dashed red curve presents the model prediction according to Eq.
%(\ref{P_n neat onset}). The solid red line was obtained by numerically
%integrating the model's equations of motion at the first threshold power with
%nonvanishing noise (Fig \ref{SM-Model}$(\QTR{rm}{a})$) and evaluating the
%spectral density.}}{\Qlb{compareSpectralDensityPower2Model}}{fig10.eps}%
%maintain-aspect-ratio TRUE;  display "ICON";  valid_file "F";
%width 3.4006in;  height 2.4574in;  depth 0pt;  original-width 7.2279in;
%original-height 5.2063in;  cropleft "0";  croptop "1";  cropright "1";
%cropbottom "0";  filename 'fig10.eps';file-properties "XNPEU";}}}%
%BeginExpansion
\begin{figure}
[ptb]
\begin{center}
\includegraphics[
height=2.4574in,
width=3.4006in
]%
{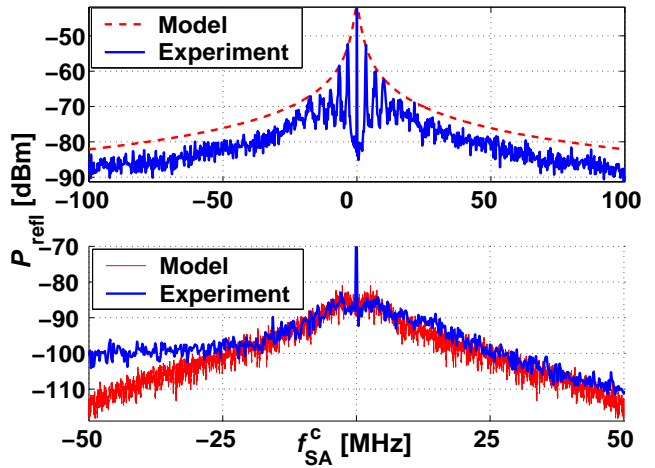}%
\caption{(Color Online) The solid blue curves in the upper and lower subplots
show typical SM experimental results in the frequency domain, obtained with
E15 and E16 respectively. The dashed red curve presents the model prediction
according to Eq. (\ref{P_n neat onset}). The solid red line was obtained by
numerically integrating the model's equations of motion at the first threshold
power with nonvanishing noise (Fig \ref{SM-Model}$(\mathrm{a})$) and
evaluating the spectral density.}%
\label{compareSpectralDensityPower2Model}%
\end{center}
\end{figure}
%EndExpansion

The spectral power density of the reflected power from\ the resonator is
predicted by Eq. (\ref{P_n neat onset}) for the case where the pump power is
slightly above the first SM power threshold, namely regular SM with a rather
low frequency occurs. This prediction is compared with typical experimental
results, obtained with E15, in the upper subplot of Fig.
\ref{compareSpectralDensityPower2Model}. The noise has a negligible influence
on the SM characteristics in that power range. On the other hand, the dynamics
of the system is governed by the noise on the edge of the SM, as is
demonstrated in Fig \ref{SM-Model}$(\mathrm{a})$. The frequency domain of the
numerical results at that region is compared with typical experimental
results, obtained with E16 in the lower subplot. Both comparisons show a good
agreement. The numerical results indicate that the model predicts a strong
noise rise near the SM power threshold. Theoretically, such noise
amplification is expected to increase when the system is approaching a
threshold of instability, where a linear theory predicts an unbounded increase
of fluctuation \cite{bifAmp_Wiesenfeld85}, which only saturates due to the
high order nonlinear terms \cite{BifAmp_Kravtsov03}. As predicted
theoretically \cite{BifAmp_Wiesenfeld86} the same mechanism generates large
signal amplification, as was indeed observed in Ref. \cite{segev06d}.

\section{Conclusions}

We report on a novel nonlinear behavior, where SM is generated in SC microwave
stripline resonators. This phenomenon is robust and occurs with all of our
devices, despite differences in geometry, and at various resonance frequencies
in each device. A theoretical model according to which the SM originates by a
thermal instability is proposed, to account for our findings. In spite of its
simplicity the model exhibits a good quantitative agreement with the
experimental results. These devices can serve as ultra-low noise amplifiers
with possible applications in the field of quantum data processing.

\begin{acknowledgments}
We thank Ron Lifshitz, Mile Cross, Oded Gottlieb, and Steven Shaw for valuable
discussions. This work was supported by the German Israel Foundation under
grant 1-2038.1114.07, the Israel Science Foundation under grant 1380021, the
Deborah Foundation, the Poznanski Foundation, and MAFAT.
\end{acknowledgments}

\bibliographystyle{apsrev}
\bibliography{Bibilography}

\end{document}